\documentstyle[preprint,pre,aps,epic,eepic,epsfig,rotate,multicol]{revtex} 
\begin{document}                                     
\pagestyle{myheadings}
\markboth{Helbing/Keltsch/Moln\'{a}r: Modelling the Evolution of Human Trail Systems}
{Helbing/Keltsch/Moln\'{a}r: Modelling the Evolution of Human Trail Systems}
\draft
\author{Dirk Helbing}
\address{II. Institute of Theoretical Physics, Pfaffenwaldring 57/III, 70550
  Stuttgart, Germany}
\author{Joachim Keltsch}
\address{Science+Computing, Hagellocher Weg 71, 72070 T\"ubingen, Germany}
\author{P\'{e}ter Moln\'{a}r}
\address{The Center of Theoretical Studies of Physical Systems,
223 James P. Brawley Drive, 
Atlanta, Georgia 30314, USA}
\title{Modelling the Evolution of Human Trail Systems} 
\maketitle
\vspace*{5mm}
{\rm Many human social phenomena, suchh as cooperation \cite{Ax,Hub,Hel3}, the
  growth of settlements \cite{Mak}, traffic dynamics \cite{Prig1,Prig2,Hel2}
  and pedestrian movement \cite{Hel2,Hend1,Hend2,Hel1}, appear to
  be accessible to mathematical descriptions that invoke
  self-organization \cite{Hak1,Pri}. 
  Here we develop a model of pedestrian motion to explore
  the evolution of trails in urban green spaces such as parks. Our aim is to
  address such questions as what the topological structures of these trail
  systems are \cite{Schenk}, 
  and whether optimal path systems can be predicted for urban
  planning. We use an `active walker' model 
  \cite{Lam1,Lam2,Schw,Lutz,Jacob,Frank} that takes into account pedestrian
  motion and orientation and the concomitant feedbacks with the surrounding
  environment. Such models have previously been applied to the study of
  complex structure formation in physical  \cite{Lam1,Lam2,Schw}, 
  chemical \cite{Lutz} and biological \cite{Jacob,Frank} systems. We
  find that our model is able to reporduce many of the observed large-scale
  spatial features of trail systems.}
\clearpage
Previous studies have shown that various observed self-organization phenomena
in pedestrian crowds can be simulated very realistically. 
This includes the emergence of lanes of uniform 
walking direction and oscillatory changes of the passing direction at
bottlenecks \cite{Hel2,Hel1}.
Another interesting collective effect of
pedestrian motion, which we have investigated very recently,
is the formation of trail systems in green areas. In many cases,
the pedestrians' desire to take the shortest way and the specific properties
of the terrain are insufficient for an explanation of the
trail characteristics. It is essential
to include the effect of human orientation. To simulate the typical
features of trail systems, we have extended the 
afore mentioned model of pedestrian motion to an active walker model
by introducing equations for environmental changes and their impact on
the chosen walking direction.
\par
First, 
we represent the ground structure at place $\vec{r}$ and time $t$
by a function $G(\vec{r},t)$ which reflects the comfort of walking.
Trails are characterized by particularly large values of $G$. 
On the one hand, at their positions $\vec{r} = \vec{r}_\alpha(t)$, 
all pedestrians $\alpha$ leave footprints on the ground 
(e.g. by trampling down some vegetation).
Their intensity is assumed to be $I(\vec{r}) [1 - G(\vec{r},t)/
G_{\rm max}(\vec{r})]$, since the clarity of a
trail is limited to a maximum value $G_{\rm max}(\vec{r})$. This causes
a saturation effect $[1 - G(\vec{r},t)/
G_{\rm max}(\vec{r})]$ of the ground's alteration by new footprints. 
On the other hand, the ground structure 
changes due to the vegetation's regeneration. This will lead to a
restoration of the natural ground conditions $G_0(\vec{r})$ with a 
certain weathering rate $1/T(\vec{r})$ which is related to the durability
$T(\vec{r})$ of trails. 
Thus, the equation of environmental changes reads
\begin{equation}
 \frac{dG(\vec{r},t)}{dt} = \frac{1}{T(\vec{r})}
 [ G_0(\vec{r}) - G(\vec{r},t)] 
 + I(\vec{r}) \left[ 1 - \frac{G(\vec{r},t)}
 {G_{\rm max}(\vec{r})} \right] \sum_\alpha 
 \, \delta (\vec{r} - \vec{r}_\alpha(t)) \, ,
\label{ground}
\end{equation}
where $\delta(\vec{r} - \vec{r}_\alpha)$ denotes Dirac's delta function
(which yields only a contribution for $\vec{r} = \vec{r}_\alpha$). 
\par
The attractiveness of a trail segment at place $\vec{r}$ 
from the perspective of place $\vec{r}_\alpha$ decreases with its distance 
$\|\vec{r} - \vec{r}_\alpha(t)\|$ and depends on the
visibility $\sigma(\vec{r}_\alpha)$. Considering this by a factor
$\exp(-\|\vec{r} - \vec{r}_\alpha\|/\sigma(\vec{r}_\alpha))$ and
taking the spatial average by integration of
the weighted ground structure over the green area, we obtain
\begin{equation}
 V_{\rm tr}(\vec{r}_\alpha,t) = \int d^2 r\, \mbox{e}^{-\|\vec{r} - 
 \vec{r}_\alpha\|/\sigma(\vec{r}_\alpha)} G(\vec{r},t) \, .
\label{potential}
\end{equation}
The trail potential $V_{\rm tr}(\vec{r}_\alpha,t)$
reflects the attractiveness of walking at place $\vec{r}_\alpha$. It
describes indirect long-range interactions 
via environmental changes, which are essential for the characteristics of the
evolving patterns \cite{Jacob}.  
\par
On a plain, homogeneous ground, the walking
direction $\vec{e}_\alpha$ of pedestrian $\alpha$ 
is determined by the direction of the next destination 
$\vec{d}_\alpha$, i.e. $\vec{e}_\alpha(\vec{r}_\alpha) = 
(\vec{d}_\alpha - \vec{r}_\alpha ) / \| \vec{d}_\alpha - \vec{r}_\alpha \|$. 
Without a destination, a pedestrian is expected to move into the
direction of the largest increase of ground attraction, which is given
by the (normalized)
gradient $\vec{\nabla}_{\!\vec{r}_\alpha} V_{\rm tr}(\vec{r}_\alpha,t)$
of the trail potential. However, since the choice of the 
walking direction $\vec{e}_\alpha$
is influenced by the destination and existing trails
at the same time, the orientation relation
\begin{equation}
 \vec{e}_\alpha(\vec{r}_\alpha,t) = \frac{\vec{d}_\alpha - \vec{r}_\alpha
 + \vec{\nabla}_{\!\vec{r}_\alpha} V_{\rm tr}(\vec{r}_\alpha,t)}
 { \| \vec{d}_\alpha - \vec{r}_\alpha
 + \vec{\nabla}_{\!\vec{r}_\alpha} V_{\rm tr}(\vec{r}_\alpha,t) \| } 
\label{orient}
\end{equation}
was taken as the arithmetic average of both effects.
Considering cases of rare interactions, the approximate
equation of motion of a
pedestrian $\alpha$ with desired velocity $v_\alpha^0$ is 
\begin{equation}
 \frac{d\vec{r}_\alpha}{dt} = v_\alpha^0 \vec{e}_\alpha(\vec{r}_\alpha,t) \, .
\label{motion}
\end{equation} 
\par
\begin{figure}[htbp]
  \begin{center}
    \leavevmode \epsfig{width=12cm, angle=0, 
      bbllx=0pt, bblly=0pt, bburx=131pt, bbury=88pt, 
      file=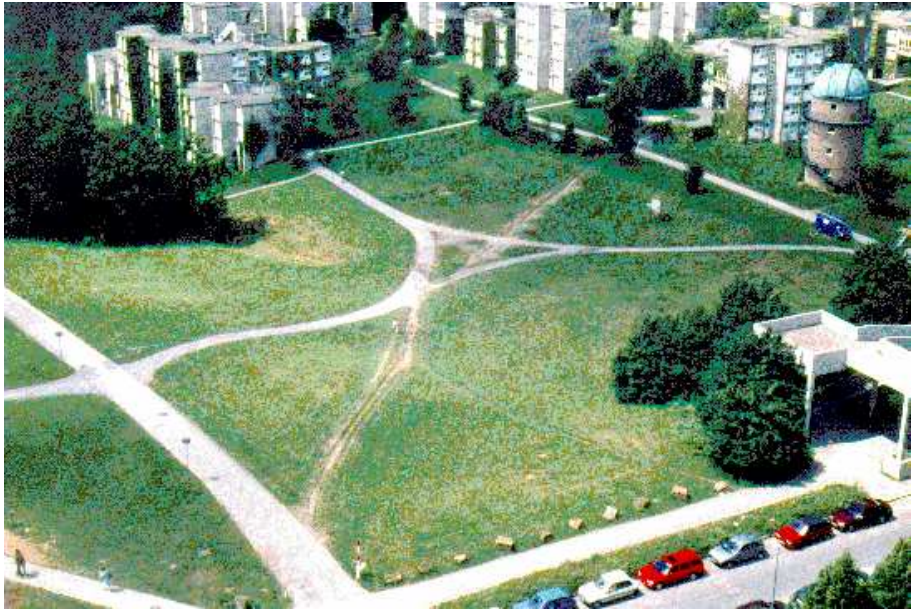}
  \end{center}
  \caption[Struktur von Trampelpfaden]
  {Between the straight, paved ways on the university campus 
  in Stuttgart-Vaihingen a trail system has evolved (center of the 
 picture). Two types of nodes are observed: 
 Intersections of two trails running in a straight line and
 junctions of two trails which smoothly merge into one trail.}
 \label{trail_vaih}
\end{figure}
\begin{figure}[htbp]
\begin{center}
\unitlength1cm
{\epsfig{width=5.2cm, angle=0, file=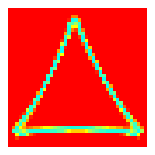}}
  \hfill
{\epsfig{width=5.2cm, angle=0, file=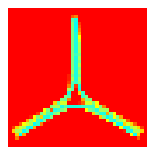}}
  \hfill
{\epsfig{width=5.2cm, angle=0, file=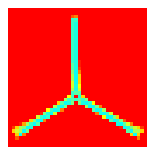}}
\end{center}
\caption[Abh\"angig von der Pfadattraktion resultierende Wegesysteme]
{The structure of the emerging trail system (light grey) 
essentially depends on the attractiveness parameter
$\kappa$. If $\kappa$ is small, a direct way system develops
(left), if $\kappa$ is large, a minimal way system is formed,
otherwise a compromise between both extremes will result (middle) which looks
similar to the trail system in the center of Figure \ref{trail_vaih}.}
\label{wegtyp}
\end{figure}

A comparison of simulation results with photographs
shows that the above described model is in good agreement with 
empirical obvervations. In particular, the evolution of the unexpected 
`island' in the middle of the trail system in Figure~\ref{trail_vaih} 
can be correctly described (Figure~\ref{wegtyp}). 
The goodness of fit of the model is quite surprising, since
it contains only two independent parameters $\kappa = I T/\sigma^2$ and
$\lambda = V^0 T/\sigma$, where $V^0$ denotes the average of the desired velocities
$v_\alpha^0$. This can be shown by scaling the model to dimensionless
equations. The parameter $\lambda$ was kept constant.
\par
\begin{figure}[htbp]
\begin{center}
    \leavevmode
    \epsfig{width=0.45\textwidth, bbllx=35pt, bblly=163pt, bburx=560pt, 
   bbury=690pt, clip=, file=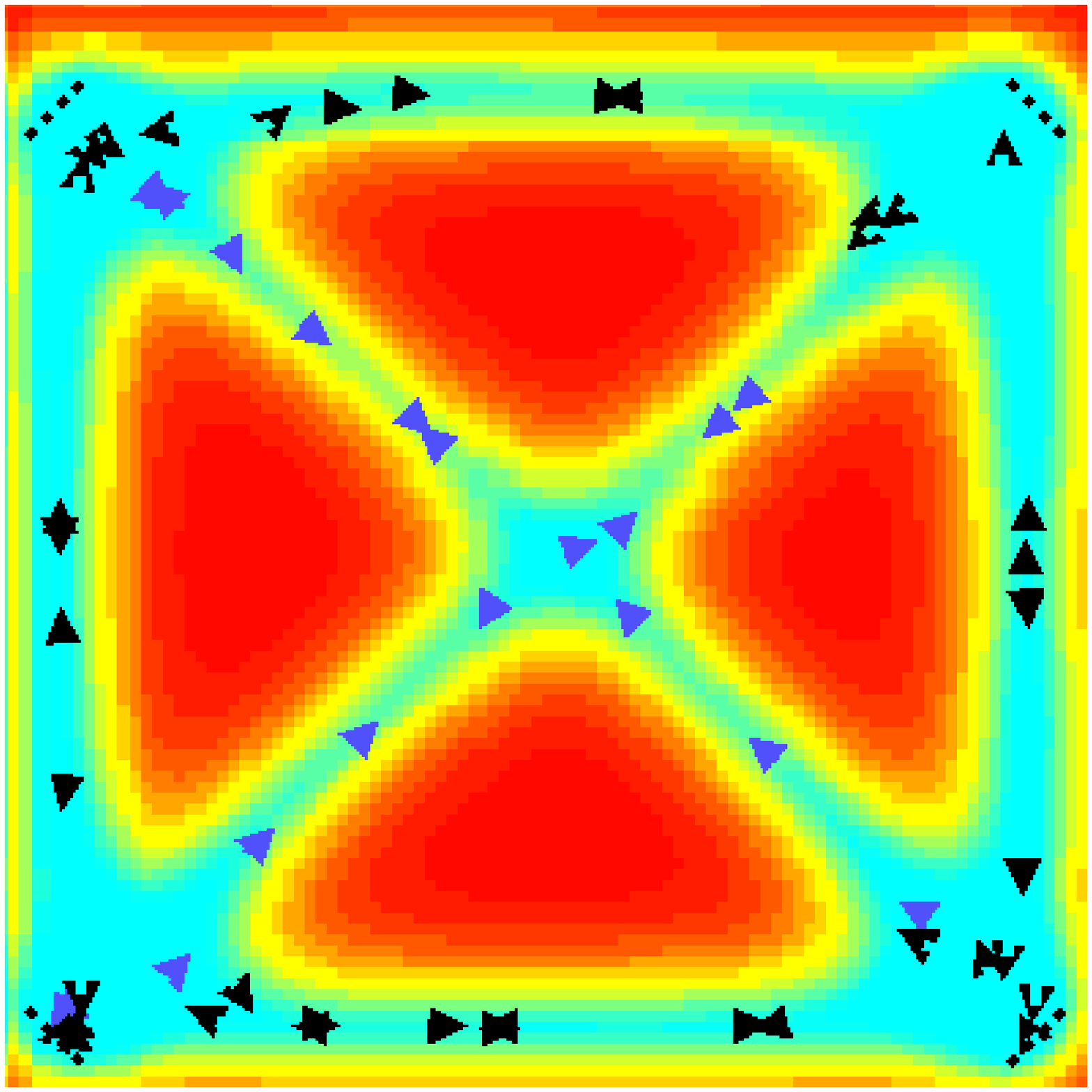} \hspace{0.5cm}
    \epsfig{width=0.45\textwidth, bbllx=35pt, bblly=163pt, bburx=560pt, 
   bbury=690pt, clip=, file=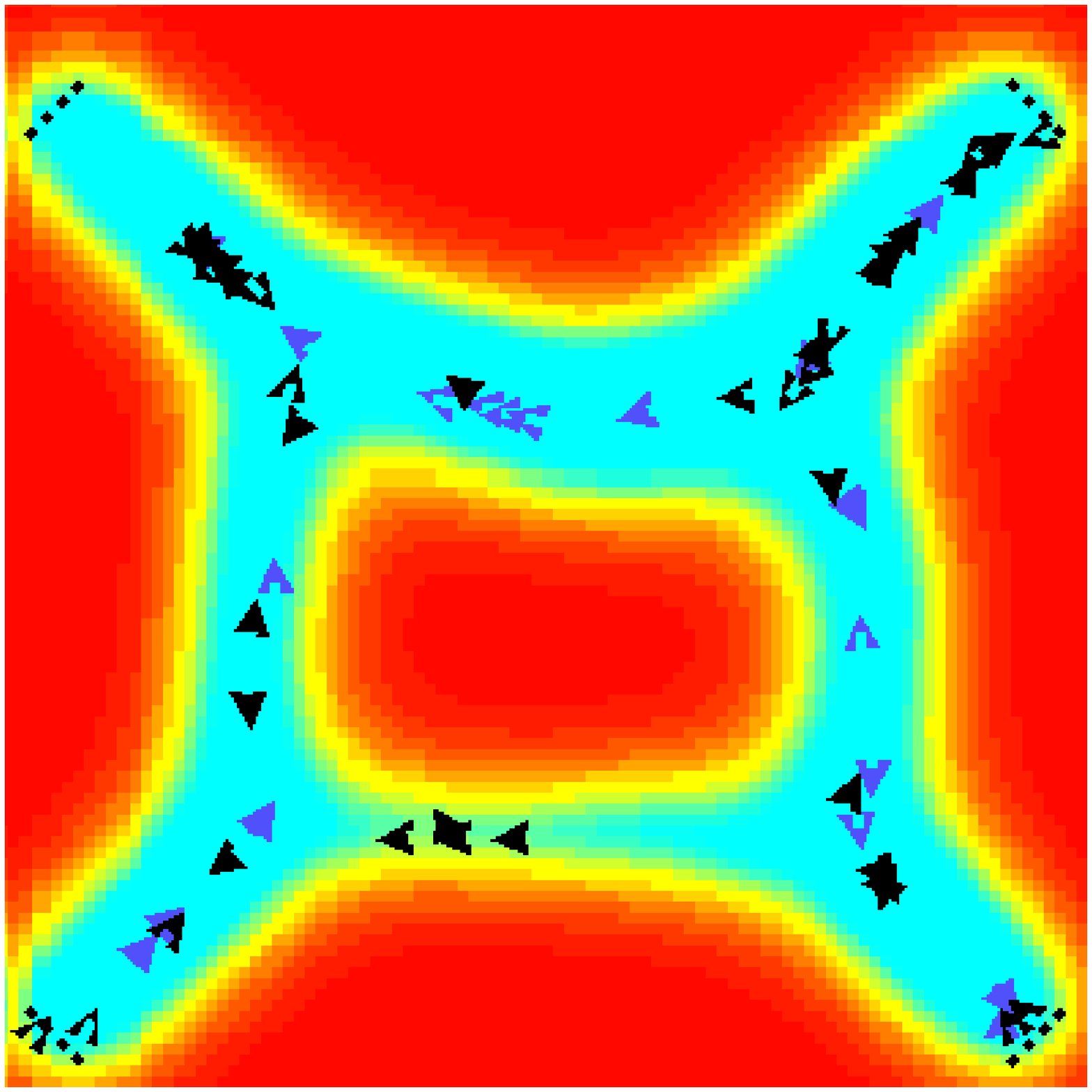} 
\end{center}
\caption[Vergleich verschiedener Wegenetztypen]
{The graphics illustrate the places and walking directions of pedestrians
by arrows. The trail potential $V_{\rm tr}(\vec{r},t)$ is represented by a color 
scale (red\,=\,small, blue\,=\,large values).
Starting with a plain ground, the structure of the trail system changes
  considerably during the simulation. Initially, pedestrians use the
direct ways (left). Since frequently used trails
become more comfortable, a bundling of trails sets in which reduces the
overall length of the trail system (right). 
The resulting way system (whose asymmetry is caused
by differences in the frequency of trail usage) could serve as a 
planning guideline. It provides a suitable compromise between
minimal construction costs and maximal comfort. Moreover, it balances
the relative detours of all walkers.}
\label{wegsys}
\end{figure}
\begin{figure}[htbp]
  \begin{center}
\unitlength1cm
\begin{picture}(12.5,11)
\put(0,0){\epsfig{height=10.75cm, 
      bbllx=177pt, bblly=350pt, bburx=320pt, bbury=720pt, angle=180, clip=,
      file=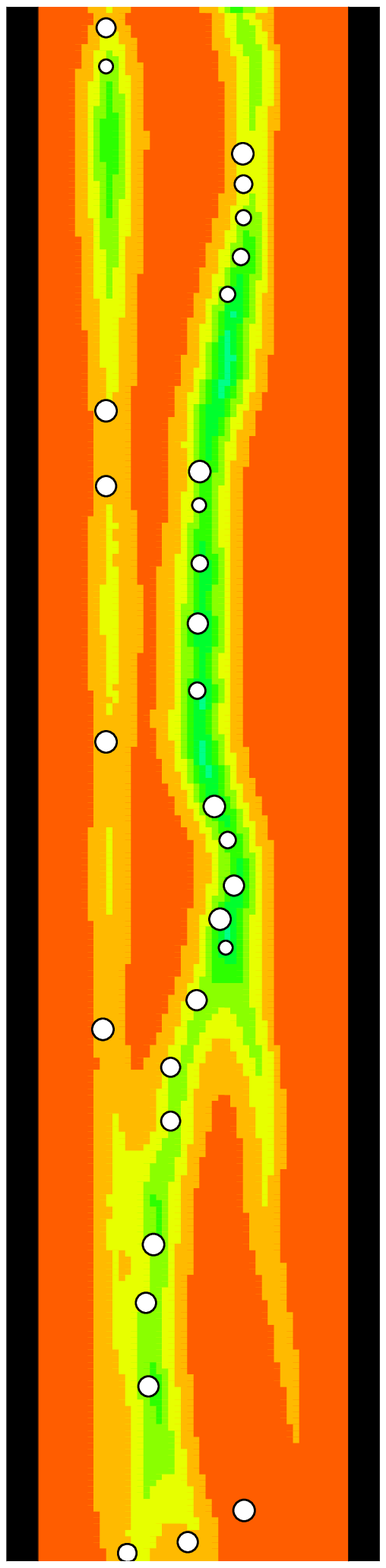}}
\put(4.8,0){\epsfig{height=11cm, angle=0, 
      bbllx=508pt, bblly=0pt, bburx=596pt, bbury=131pt, 
      file=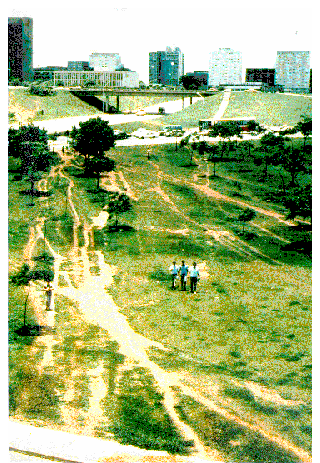}}
\end{picture}
\end{center}
\caption[Spurbildung eines Fu{\ss}g\"angerstroms auf einer schmalen
  Rasenfl\"ache]
 {When pedestrians leave footprints on the ground, trails will develop,
 and only parts of the ground are used for walking (in contrast to
 paved areas). The similarity between the simulation result (left)
 and the trail system on the university
 campus of Brasilia (right, reproduction by kind permission of
 Klaus Humpert) is obvious.}
\label{tramp}
\end{figure}

Our simulations base on a discretization of the considered area in
small quadratic elements of equal size, which converts the integral
(\ref{potential}) into a sum. Temporal and spatial derivatives are approximated
by difference quotients. The presented examples 
begin with plain, homogeneous ground. All pedestrians have
their own destinations and entry points, from which they start at a randomly
chosen point in time. In Figure \ref{wegtyp} (Figure \ref{wegsys}) pedestrians
move between all possible pairs of three (four) fixed places. 
While in Figure \ref{tramp} the entry points and destinations
are distributed over the small ends of the ground.
\par
At the beginning, pedestrians take the direct ways to their respective
destinations. However, after some time they begin to use already existing
trails, since this is more comfortable than to clear new ways. 
By this, a kind of selection process 
\cite{Eig,Ebel,Schw} between trails sets in:
Frequently used trails are more attractive 
than others. For this reason they are chosen very often, and the resulting
reinforcement makes them even more attractive.
However, the weathering effect destroys rarely used trails and
limits the maximum length of the way system 
which can be supported by a certain rate 
of trail usage. As a consequence, 
the trails begin to bundle, especially where different 
trails meet or intersect. This explains, why pedestrians with different 
destinations use and produce common parts of the trail system
(Figures \ref{wegtyp} and \ref{wegsys}). 
\par
A direct way system (which provides the shortest connections, but
covers a lot of space) only develops if all ways are
almost equally comfortable. If the advantage $\kappa$ of 
using existing trails is large, the final trail system is a
minimal way system (which is the shortest way system that connects all
entry points and destinations). For realistic values of
$\kappa$, the evolution of the trail
system stops before this state is reached (Figure~\ref{wegtyp}).
Thus, $\kappa$ is related to the average relative detour
of the walkers. We conjecture that
the resulting way system is the shortest one which 
is compatible with a certain accepted relative detour.
In this sense, it yields an optimal compromise between 
convencience and shortness.
\par 
Therefore, we suggest to use the above model as a tool for
urban planners and landscape gardeners, who have the dilemma to 
build most comfortable way systems at minimal construction costs.
For planning purposes one needs to know the entry points and destinations
within the considered area and the rates of usage of their connections.
If necessary, these can be estimated by trip chaining models \cite{Timmer},
which are also needed in cases of complex lines of access and sight. 
The effects of the physical terrain and already existing
ways can be taken into account by the function $G_0(\vec{r})$.
By varying the model parameter $\kappa$, the overall length of the resulting
trail system can be influenced (Figures \ref{wegtyp} and \ref{wegsys}). 
In the same way,
one can check its structural stability. Presently, we are evaluating typical
parameter values of $\lambda$ and $\kappa$ by comparison of simulation
results with real pedestrian flows which are reconstructed from video
films by image processing. These values shall be used 
for designing convenient way systems in
residential areas, parks, and recreation areas by means of computer
simulations (Figure~\ref{wegsys}).
We expect that such way systems will actually be accepted, since they take into account the
route choice habits of pedestrians.
\par
In summary, the presented active walker model is able to describe the
self-organization and the typical structural properties of human trail systems.
It will be interesting to 
relate our model to the work on space syntax \cite{Hillier}.
Repulsive interactions between pedestrians can be taken into account by
generalizing equation (\ref{motion}) in accordance with
the social force model of pedestrian motion \cite{Hel2,Hel1}. However, these
are only relevant in cases of frequent pedestrian interactions, in which they
lead to broader trails.

\section*{Acknowledgments}

The authors would like to thank Frank Schweitzer for many stimulating
discussions. Wolfgang Weidlich and Martin Treiber were helpful 
in reviewing the manuscript.

\clearpage
\clearpage
\end{document}